# Little-Parks effect in single YBaCuO sub-micron rings

Franco Carillo[1], Gianpaolo Papari[2], Daniela Stornaiuolo[2,3], Detlef Born[1], Domenico Montemurro[2,1], Pasqualantonio Pingue[1], Fabio Beltram[1], Francesco Tafuri[1,3].

[1] NEST CNR-INFM and Scuola Normale Superiore, Piazza San Silvestro 12, I-56127, Pisa, Italy
[2] Dipartimento Scienze Fisiche and SPIN INFM-CNR, Università di Napoli Federico II, Napoli, Italy
[3] Dipartimento di Ingegneria dell'Informazione and SPIN INFM-CNR, Seconda Università di Napoli, Aversa, I-81031 Caserta, Italy

The properties of single submicron high-temperature superconductor (HTS) rings are investigated. The Little-Parks effect is observed and is accompanied by an anomalous behavior of the magnetic dependence of the resistance, which we ascribe to non-uniform vorticity (superfluid angular momentum) within the ring arms. This effect is linked to the peculiar HTS-relationship between the values of the coherence length and the London penetration depth.

# I-Introduction

Rings and, more generally, multiply connected systems are powerful tools to investigate purely quantum phenomena such as gauge invariance and quantum interference in superconducting [1] and normal-state systems [2]. In particular Aharonov-Bohm [3] and Little-Parks [4-6] effects represent paradigmatic examples of coherent quantum phenomena occurring in simple loops for the case of single particles and of the collective superconducting state, respectively. In the case of mesoscopic structures made of low critical-temperature superconductors (LTS), devices with characteristic dimensions comparable or lower than the coherence length ($\xi_0$) are today accessible and allowed a number of exciting experiments on superconductivity at the nano-scale [5-8]. Similar scaling regimes cannot be achieved in the case of high critical-temperature superconductors (HTS) because of the very small value of the coherence length in perfectly doped HTS ($\xi_0 \approx 1.5$ nm). Nevertheless currently available linewidths of the order of few tens of nanometers in HTS open the way to the experimental study of structures with length scales comparable or smaller than the London penetration length ($\lambda_L$), or the typical grain size in epitaxial thin films, or the possible stripe charge order correlation length (40nm) [9-11]. Indeed, rings on the deep submicron scale are expected to shed light on various issues related to the HTS pairing mechanism [9-13] and to the peculiar vortex states produced by the d-wave order-parameter symmetry (OPS) [12-18]. Also, it was recently claimed that since in a d-wave system robust superconductivity along specific directions can coexist with low-energy quasi-particles, a crossover between the typical h/2e superconducting and h/e quasi-particle flux-periodicity [12,14,15,18] should be observed in mesoscopic loops of length scales even larger than $\xi_0$.

In the present work we report h/2e Little-Parks flux periodicity in nanoscale HTS rings. When compared with analogous experiments performed on LTS loops with a similar ratio between internal and external radii [19, 20] or on arrays of micron-sized HTS holes [21], our results provide clear evidence of a non-uniform vorticity of the order parameter in the rings. We shall discuss this "concentric-vortex" structure which results from the very short coherence length and a London penetration depth $\lambda_L$ which is, in turn, larger than the width of the ring arms. It is worth mentioning that since $\xi$ is of the order of few nm, our findings are relevant even to ring radii smaller than the ones studied here, down to the nm scale.

Isothermal measurements of resistance vs. magnetic field (R(H)) of submicron rings patterned on YBCO c-axis thin films were performed on several different samples. By measuring R(H) at temperatures close to the critical temperature ($T_c$) in a mesoscopic loop [5,6] it is possible to determine its normal-to-superconducting phase boundary, in full analogy with the experiments reported by Little and Parks [4] on cylinders to evidence fluxoid quantization [1]. As explicitly shown in ref. 5, in structures of characteristic size comparable to ξ, the phase boundary strongly depends on the specific shape of the structure.

In HTS the detection of $T_c$ vs. $H$ oscillations can be very challenging since their amplitude ($\Delta T_c$) is expected to scale with $(\xi_0/r)^2$. In one of the few successful attempts a maximum $\Delta T_c \approx 40$ mK was observed in a 50×50 array of 1µm×1µm holes in YBaCuO c-axis films [21]. In our rings with a radius of a few hundred nanometers, $\Delta T_c$ oscillations show an amplitude of about 500 mK on single YBaCuO loops. In our single-ring configuration higher spurious harmonics in the magnetoresistance are not introduced as opposed to arrays of rings. In these last systems the magnetoresistance is the convolution of the fundamental period, which corresponds to a flux quantum in a single cell, but also of all the periods corresponding to a flux quantum in two or several cells of the array. Consequently in an array, it is difficult to distinguish these contributions from intrinsic effects with integer fractions of h/2e in a single loop [22].

## II-Experiment

We shall report on measurements performed on four YBaCuO rings having nominally identical shape (internal radius $r_i = 200$nm, external radius $r_o = 500$nm), but different critical temperatures ranging from 74K to 19K. These samples allow us to cover a large part of the superconducting phase diagram in the under-doped regime. At lower $T_c$ interference due to quasiparticles is expected to play a larger role also owing to the longer thermal length. To the best of our knowledge single-particle Aharanov-Bohm interference was recorded up to the maximum temperature of 20K on semiconductor rings with similar size [23].

We fabricated our nanostructures starting from 50 nm-thick (001)-oriented YBCO films grown on YSZ and $SrTiO_3$ substrates. The typical value of the surface roughness was lower than 3 nm. A 20 nm gold layer was deposited *in situ* to protect YBCO films during all following fabrication steps. In order to fabricate the nanostructures we deposited on top of the YBCO/Au bi-layer a 50 nm-thick Ti mask that was patterned by electron-beam lithography and lift-off. The exposed YBCO/Au bilayer was etched by ion-beam etching (IBE) while the sample was cooled down at -140 °C in order to minimize oxygen loss from YBCO films. Subsequently the Ti mask was removed by a highly-diluted HF solution (1:20) and the gold layer by a short and low-energy IBE. Figure 1c reports a scanning electron microscope picture of one of the nanostructures (D1).

The critical temperature of a device patterned using this procedure depends on the linewidth and on etching conditions. The more energetic is the dry etching and the smaller are the linewidths, the more depressed are the superconducting properties of nanostructures. This makes it possible to tune the ring doping level. We measured four rings, with an average radius of 350 nm and branch width ranging from 270 to 300nm. For all devices the transition temperature of the larger YBCO areas, i.e. wiring and pads, corresponds to that of the unpatterned film (86K). All samples showed the expected tail in the resistance vs. temperature curve associated to the superconducting transition of the submicron part of the device (see Fig. 1a for device D1). In Figure 1b we show R(T) curves at zero magnetic field and at 12 T close to $T_c$ for a similar ring with a lower doping. In experiments on single crystals and thin films, the exponential growth of the resistivity in an extended temperature range above the zero resistance state [24] has been associated to a vortex liquid state (VLS) [25,26] in the complex magnetic phase diagram. Even if the conditions of occurrence of VLS in HTS nanostructures go beyond the aims of the present work, the R(T) curve, measured at 12 T (which is the maximum field we can apply in our set-up) and shown in Figure 1b close to $T_C$, gives relevant indications. Its characteristic shape accompanied by a reduction of $T_C$ of about 25 K (of the order of 40% of the original $T_C$), suggests a mechanism based on VLS for the onset of resistance also in our nanostructures, where the width is comparable to the London penetration depth and much smaller than the Pearl length.

Measurements were performed in liquid helium using a variable-temperature probe. Thermal stability was ensured by a heater connected to a PID controller. Current and voltage leads were filtered by RC filters at room temperature. A second filtering stage consisted of RC plus copper-powder filters thermally anchored at 4.2K. Shielding from spurious external magnetic fields was provided by a combination of nested cryoperm, Pb and Nb foils all placed in the measurement dewar and immersed in liquid He. For all the measurements reported here we consider the differential resistance (dR = dV/dI) at zero bias as function of temperature and magnetic field. dR is measured by biasing the device with a small ac current (typically 300 nA) at about 11 Hz and detecting the corresponding ac voltage with a lock-in amplifier.

Following the standard procedure of LP experiments on LTS samples, we measured the magnetoconductance at several temperatures above $T_{C0}$ defined as the temperature at which the voltage drop is zero within the error of the instrument. From the magnetoconductance and the R(T) curves we calculated $\Delta T_C = \Delta R \left( \frac{dR}{dT} \right)^{-1}$. In all devices $\Delta Tc$ oscillations show a periodicity h/2e as function of the flux enclosed by the ring average area $A = \pi r^2_{Avg}$, as shown in Fig. 1 c for D1. For our geometry this area matches with theoretical predictions for annulus of arbitrary $r_i/r_o$ ratio [27]. Consistently with similar measurements on the LP effect [4-6, 21, 28], the periodic dependence of $\Delta Tc$ is superimposed on a parabolic background determined by the contribution to magnetoconductance of single wires [28]. The amplitude of $T_C$ oscillations depended on the temperature at which the R(H) was measured. The maximum value of $\Delta T_C = 600$ mK was recorded at 73K, and far exceeds the value predicted by theory for clean superconductors: $\Delta T_C = 0.14 T_C (\xi_0/r)^2 \sim 1$ mK, assuming a zero-temperature coherence length $\xi_0 = 1.5$nm. This large discrepancy is consistent with previous experiments in HTS [21] and conventional systems [28] and has been linked to a smooth R(T) since early times [28]. Current leads connected to the rings are quite larger than the arms of the ring (an example is given in Fig. 1c). This has guaranteed that for the presented experiments we have never been in a situation for which the leads were normal and the ring superconducting, differently from other configurations studied in literature (Ref. 29). The results in Ref. 29 suggest that in our experiment possible more subtle processes induced by non equilibrium effects produced by normal leads are negligible with respect with the shielding current phenomena in the loop (i.e. Little-Parks effect).

The magnetic field behavior of severely under-doped rings with $T_c$ 24 K and 19 K is substantially identical to those with larger doping. In particular we have found that, within our sensitivity and noise level, the oscillatory signal of magnetoresistance quickly disappears above $T_c$ (typically in a temperature range $\Delta T/T_c$ less than 5%). The absence of an oscillatory response does not necessarily argue against the scenario of Cooper pairs surviving at temperatures larger than $T_c$ in the entire pseudo-gap region [30,31]. Nevertheless our results might signal an upper bound on the length scale for coherent phenomena above $T_c$. A more specific theoretical framework able to take into account possible effects imposed by 'nano-sizes' would be necessary for a more complete asset.

The behavior at larger magnetic fields (300mT) shows a few unexpected features when compared to existing data on LTS [4-6, 28]. Data for D2 are shown in Fig. 2a (the parabolic background was subtracted). Oscillations have an amplitude modulation with at least three nodes before completely disappearing at high fields. This beating pattern results from the mix of two or more frequency components that are related to the inverse of the magnetic field. Fast Fourier transform (FFT) of the pattern shown in Fig. 2b yields all frequency components in the data. This procedure was successfully used in Ref. 32 to detect the characteristic periodicity arising from fluxoid quantization in two distinct concentric aluminum rings. In our work, apart from the expected peak matching the fundamental periodicity $\Delta H = \Phi_0/\pi r^2_{avg}$, we found additional peaks. They occur at the same position for all temperatures. Each FFT peak indicates the existence of a characteristic radius associated with flux quantization. Such a multi-frequency behavior is made possible by the size of the ring in relation to the characteristic lengths of HTS. The "multiple-peak frequency" behavior must be ascribed to the finite width (W) of the ring arms which is here much larger than the superconductor coherence length.

In LTS systems, $T_c$ oscillations were detected in hollow cylinders and rings with walls or arms thinner than $\xi$. In this case the order parameter can be considered constant along the radial direction and depends only on the azimuthal angle. In the r >> ξ regime, $T_c$ and other thermodynamic properties, e.g. the magnetization, are strictly h/2e periodic. In HTS systems it is possible to have access to a regime where the radius of the ring is small enough to yield sizable LP oscillations, while the width of the arms still supports a significant variation of the order parameter along the radial direction. This sustains a discrete number of concentric independent domains where supercurrent density is different from zero.

## III-Discussion

For a superconducting annulus close to $T_c$ with arbitrarily wide arms, Ginzburg-Landau equations can be used to determine the order parameter (ψ) which, for the cylindrical symmetry of the sample, can be written as $\psi_L(\mathbf{r}) = f_L(\rho)e^{iL\theta}$, where θ and ρ are cylindrical coordinates (origin at the center of the ring). L is called *winding number* or *vorticity* and can be thought of as the angular momentum of the superfluid density. For W << ξ, $f_L(\rho)$ is a constant, while in our case, W >> ξ, $f_L(\rho)$ varies with ρ.

In our samples the lateral penetration depth exceeds W, so that the magnetic field in the structure can be taken equal to the external magnetic field. For the nucleation of superconductivity linearized GL equations can be used [1, 5, 6] and the most general solution is a linear combination of states with different winding number $\psi(r) = \sum a_L \psi_L(r)$ [33]. Each $\psi_L(\mathbf{r})$ is characterized by a different average radius and consequently a different periodicity in H. Coefficients $a_L$ must be chosen by minimizing the free energy of the system and maximizing the flatness of ψ (see Ref. 33).

If the external radius $r_o > \pi\xi$, $f_L(\rho)$ has nodes along the radial direction (between the minimum and maximum radius) with an overall spacing ≈ πξ [34]. Consecutive nodes mark concentric domains in which the order parameter and the supercurrent are different from zero [34].

A stable solution with lower free energy may take place in a configuration with an order parameter having different vorticity in two domains of a ring separated by a zero-current line rather than the configuration with uniform vorticity [35]. In this case the total free energy of the ring will be a sum of contributions from domains with a different periodicity in H.

On the basis of these approaches [34, 35], we associate each of the peaks in the FFT to the effective radius of one of the elements of a set of concentric current loops populating the ring, each labeled by different superfluid momentum, as schematized in Fig. 3c, where the FFT is reported in polar coordinates. R(H) data reported in Fig. 2-a were divided in two groups: from 0 to 90 mT and from 90 to 250 mT: Fig. 3a shows the corresponding FFTs (solid and dashed line, respectively). From each of the marked peaks in Fig. 3a ($g_n$) we can calculate a radius value $r_n = (\Phi_0 g_n/\pi)^{0.5}$ that is plotted in Fig. 3b as function of the order of the peak. A linear fit of the data yields the average radius spacing $\Delta r$ at both higher (21 ± 3nm) and lower (28 ± 2nm) magnetic field values. From this we can estimate the GL parameter $\xi = \Delta r/\pi = 12 \pm 1$ nm. Finally from this $\xi$ value, by using the expression for a clean superconductor $\xi_0 = \xi(1-T/T_c)^{-0.5}$ with T = 70 K and $T_c$ = 74 K, we obtain $\xi_0 = 1.5 \pm 1$ nm.

The shape of the two FFTs in Fig. 3a is rather similar but the one relative to the larger magnetic fields appears somewhat compressed toward zero frequencies. This behavior is in agreement with the theoretical findings of Ref. 36. For a solution of GL equation with fixed vorticity L, the zero current line monotonously shrinks toward the internal radius of the ring with increasing magnetic field.

The multiple-peak structure observed in FFT and the shift of the peaks to lower frequency at higher magnetic field closely resemble what experimentally observed and theoretically expected in Aharonov-Bohm multimode rings in the two-dimensional electron gas [37,38]. We argue that this striking similarity holds because linearized GL equations, which model our system, are formally identical to the Schrodinger equation which describes single-particle states in a normal ring [1]. The main argument against the hypothesis of a role of quasiparticles interference is the absence of h/e periodicity [12]. Quantitative analyses also rule out explanations of multi-peaked FFT in terms of quasiparticle states with different angular momenta (and radii), as modeled in Ref. 38 for semiconductor multimode rings. In fact, in this case, the difference between two average radii associated with single particle states would be of the order $\pi\lambda_F \sim 0.3$ nm, where $\lambda_F$ is the Fermi wavelength in YBCO (~0.1nm).

## IV-Conclusions

In conclusion we performed LP experiment on several differently-doped submicron YBCO rings. Results indicate a multi-period dependence on H of the free energy close to $T_c$, consistent with a non-uniform vorticity in the ring. This is a quantum effect due to mesoscopic confinement on HTS nanostructures. We believe these findings can provide useful guidelines for the design of nanoscale experiments targeting the investigation of fundamental properties of HTS.

We acknowledge helpful discussions with J.R. Kirtley, F. Peeters, V. Moshchalkov. This work has been supported by the MIUR-FIRB RBIN06JB4C and EC-STREP "MIDAS-Macroscopic Interference Devices for Atomic and Solid State Physics: Quantum Control of Supercurrents" projects

References:

[1] M. Tinkham, *Introduction to Superconductivity* (McGraw-Hill, New York, 1975)

[2] Y. Imry, Introduction to Mesoscopic Physics, Oxford University Press, 1997

[3] Y. Aharonov, and D. Bohm, Phys. Rev. 115, 485 (1959)

[4] W.A. Little and R.D. Parks, Phys. Rev. Lett. **9**, 9 (1962)

[5] V. V. Moshchalkov, L. Gielen, C. Strunk, R. Jonckheere, X. Qiu, C. Van Haesendonck, Y. Bruynseraede, Nature 373, 319 (1995)

[6] V. V. Moshchalkov, L. Gielen, M. Dhallè, C. Van Haesendonck, Y. Bruynseraede, Nature 361, 617 (1993)

[7] A.K.Geim,, S.V. Dubonos, J.G.S.Lok, , M. Henini, , J.C. Maan, Nature, 396, 144 (1998); A.K.Geim, I.V.Grigorieva, , S.V. Dubonos, , J.G.S.Lok, , J.C.Maan, A.E.Filippov, , F.M.Peeters, Nature 390 , 259 (1997)

[8] Y. Liu, Yu. Zadorozhny, M. M. Rosario, B. Y. Rock, P. T. Carrigan, and H. Wang, Science 294, 2332 (2001)

[9] P. Mohanty, J.Y.T. Wei, V. Ananth, P. Morales, W. Skocpol, Physica C 408–410, 666 (2004)

[10] E.W. Carlson, V.J. Emery, S.A. Kivelson, D. Orgad, in: K.H. Bennemann,, J.B. Ketterson (Eds.), The Physics of Conventional and Unconventional Superconductors, Springer-Verlag, 2002, cond-mat/0206217

[11] J. A. Bonetti, D. S. Caplan, D. J. Van Harlingen, and M. B. Weissman Phys. Rev. Lett. 93, 087002 (2004), H. A. Mook, P. Dai, F. Dogan, Phys. Rev. Lett. 88, 097004 (2002).

[12] F. Loder, A. P. Kampf, T. Kopp, J. Mannhart, C. W. Schneider & Y. S. Barash, Nature Physics 4, 112 - 115 (2008). F. Loder, A. P. Kampf, and T. Kopp Phys. Rev. B **78** 174526 (2008); Y. S. Barash, Phys. Rev. Lett. 100, 177003 (2008)

[13] C.C. Tsuei and J. R. Kirtley, Rev. Mod. Phys. 72, 969 (2000)

[14] Victor Vakaryuk Physical Review Letters **101** 167002 (2008)

[15] V. Juricic, I. F. Herbut, and Z. Tesanovic, Phys. Rev. Lett. 100, 187006 (2008)

[16] J.C. Wynn, D.A Bonn, B.W. Gardner, Yu-Ju Lin, R. Liang, W.N. Hardy, J.R. Kirtley and K.A. Moler, Phys. Rev. Lett. 87, 197002 (2001); D. A. Bonn, Janice C. Wynn, Brian W. Gardner, Yu-Ju Lin, Ruixing Liang, W. N. Hardy, J. R. Kirtley & K. A. Moler, Nature **414** pp887-889 (2001).

[17] T. Senthil and M.P.A. Fisher, Phys. Rev. Lett. 86, 292 (2001)

[18] T.C. Wei and P.M. Goldbart, Phys. Rev. B 77, 224512 (2008)

[19] Mathieu Morelle, Dusan S. Golubovic, and Victor V. Moshchalkov, Physical Review B **70** 144528 (2004)

[20] V. Bruyndoncx, L. Van Look, M. Verschuere, and V. V. Moshchalkov, Phys. Rev. B 60, 10468 (1999)

[21] P.L. Gammel, P.A. Polakos, C.E. Rice, L.R. Harriott, and D.J. Bishop, Phys. Rev. B **41**, 2593 (1990).

[22] J. Wei, P. Cadden-Zimansky, and V. Chandrasekhar Appl. Phys. Lett. 92, 102502 (2008), Yu. Zadorozhny and Y. Liu, Europhys. Lett. 55, 712 (2001).

[23] F. Carillo, G. Biasiol, D. Frustaglia, F. Giazzotto, L. Sorba and F. Beltram Physica E **32**, 53 (2006)

[24] T. T. M. Palstra, B. Batlogg, L. F. Schneemeyer, and J. V. Waszczak,. Phys. Rev. Lett. 61 1662 (1988). R. H. Koch, V. Foglietti, W. J. Gallagher, G. Koren, A. Gupta, and M. P. A. Fisher, Phys. Rev. Lett. 63 1511 (1989): H. Safar, L.P. Gammel, D.A. Huse, D.J. Bishop, J.P. Rice and D.M. Ginsberg, Phys. Rev. Lett. 69, 824 (1992); W.K. Kwok, S. Fleshler, U. Welp, V.M. Vinokur, J. Downey, G.W. Crabtree and M.M. Miller, Phys. Rev. Lett. 69, 3370 (1992)

[25] P.L. Gammel, L.F. Schneemeyer, J.V. Waszczak and D.J. Bishop, Phys. Rev. Lett. 61, 1666 (1988); D. J. Bishop, P. L. Gammel, D. A. Huse and C. A. Murray, Science 255, 165 (1992); P.L. Gammel, Nature 411, 434 (2001)

[26] M.P.A Fisher, Phys. Rev. Lett. 62, 1415 (1989); D. R. Nelson and H. S. Seung, Phys. Rev. B 39, 9153 (1989), D. S. Fisher, M. P. A. Fisher, David A. Huse,. Phys. Rev. B 43, 130 (1991).

[27] V. G. Kogan, J R. Clem, R. G. Mints, Phys. Rev. B 69, 064516 (2004)

[28] R. P. Groff and R. D. Parks, Phys. Rev. **176**, 567 (1968)

[29] D. Y. Vodolazov, D. S. Golubovic, F. M. Peeters, and V. V. Moshchalkov Phys. Rev. B **76**, 134505 (2007)

[30] For a review, see T. Timusk and B. Stratt, *Rep. Prog. Phys.* **62**, 61 (1999), Patrick A. Lee, Naota Nagaosa, and Xiao-Gang Wen, *Rev. Mod. Phys.* **78**, 17 (2006).

[31] Y. Wang, L. Li, and N. P. Ong, Phys. Rev. B 73, 024510 (2006)

[32] M. Morelle, V. Bruyndoncx, R. Jonckheere, and V. V. Moshchalkov, Phys. Rev. B 64, 064516 (2001)

[33] V.V. Moshchalkov, M. Dhallé, and Y. Bruynseraede, Physica C **207**, 307 (1993)

[34] G. Stenuit, J. Govaerts, D. Bertrand, and O. van der Aa, Physica C **332**, 277 (2000)

[35] H. Zhao, V. M. Fomin, J. T. Devreese, and V. V. Moshchalkov, Solid State Commun. **125**, 59 (2003).

[36] S. V. Yampolskii, F. M. Peeters, B. J. Baelus, and H. J. Fink, Phys. Rev. B **64**, 052504 (2001).

[37] J. Liu, W. X. Gao, K. Ismail, K. Y. Lee, J. M. Hong, and S. Washburn, Phys. Rev. B **48**, 15148 (1993).

[38] W.-C. Tan and J. C. Inkson, Phys. Rev. B **53**, 6947 (1996). Tan, W.-C. & Inkson, J. C. Semicond. Sci. Technol. 11, 1635–1641 (1996).

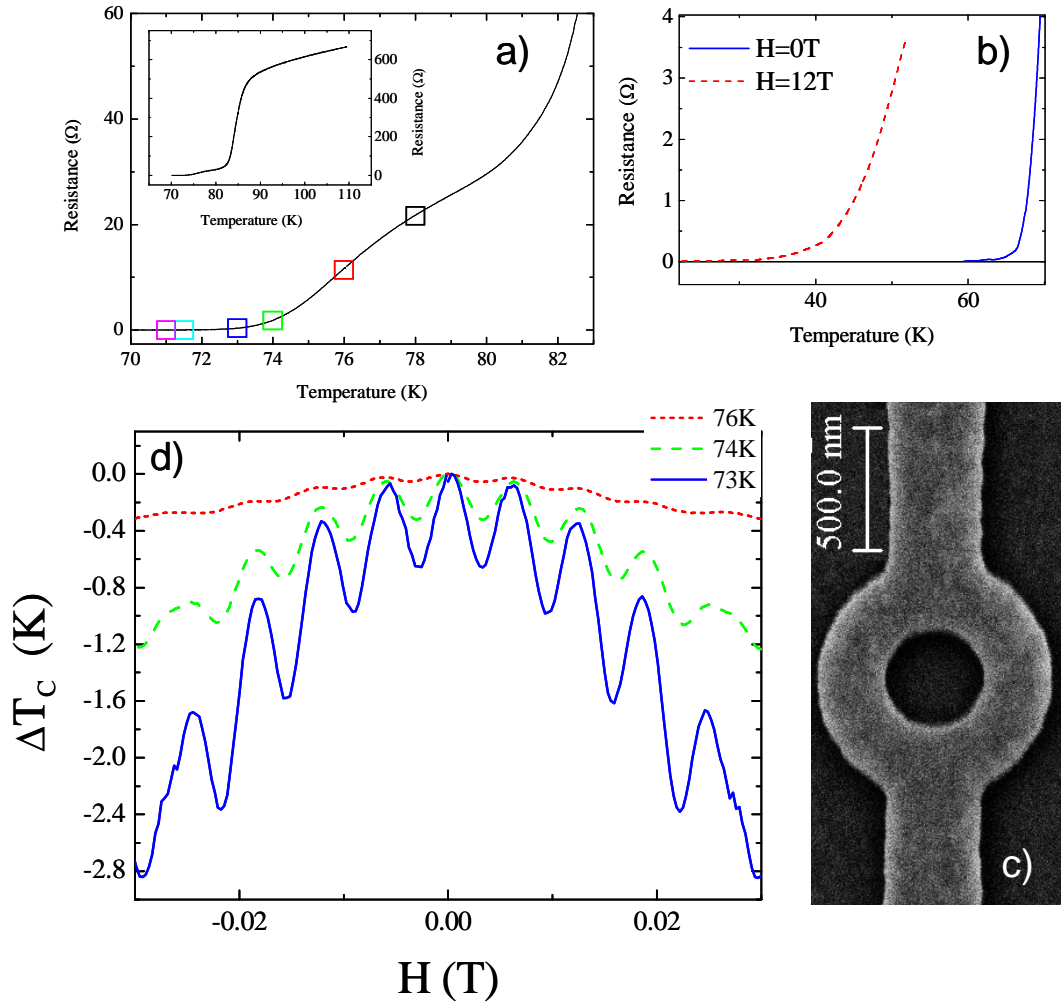

Figure 1 (Color online) (a) Detail of the Resistance vs. Temperature curve in proximity of the critical temperature Tc and on a larger scale (inset) for ring D1. In (b) R(T) curves measured at H = 0 T and H = 12 T, in an underdoped ring nominally identical to D1, are reported in proximity of  $T_c$. In (c) scanning electron micrograph of ring (D1). In (d) oscillations of  $\Delta Tc$ derived by measurements of the magneto-conductance (see text) on the same device, are shown at different temperatures, providing evidence of Little-Parks effect.

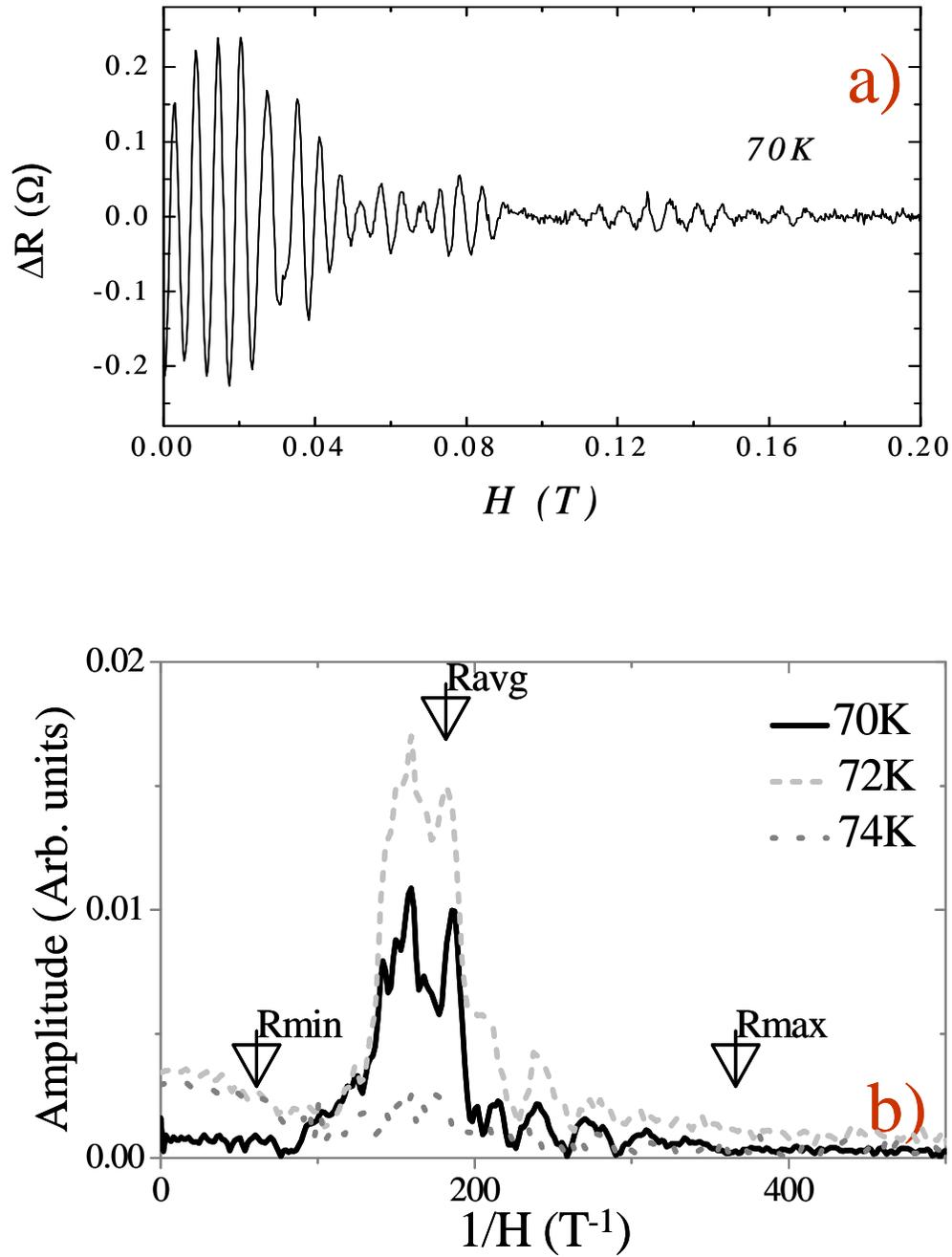

Figure 2 (Color online) a) Resistance oscillations of a ring whith $R_{max}$=445nm and $R_{min}$=205nm. (D2) b) FFT of R(H) data taken at three temperatures. Black arrows indicate the frequencies corresponding to a single flux quantum in a ring with radius $R_{min}$, $R_{avg}$, $R_{max}$ respectively.

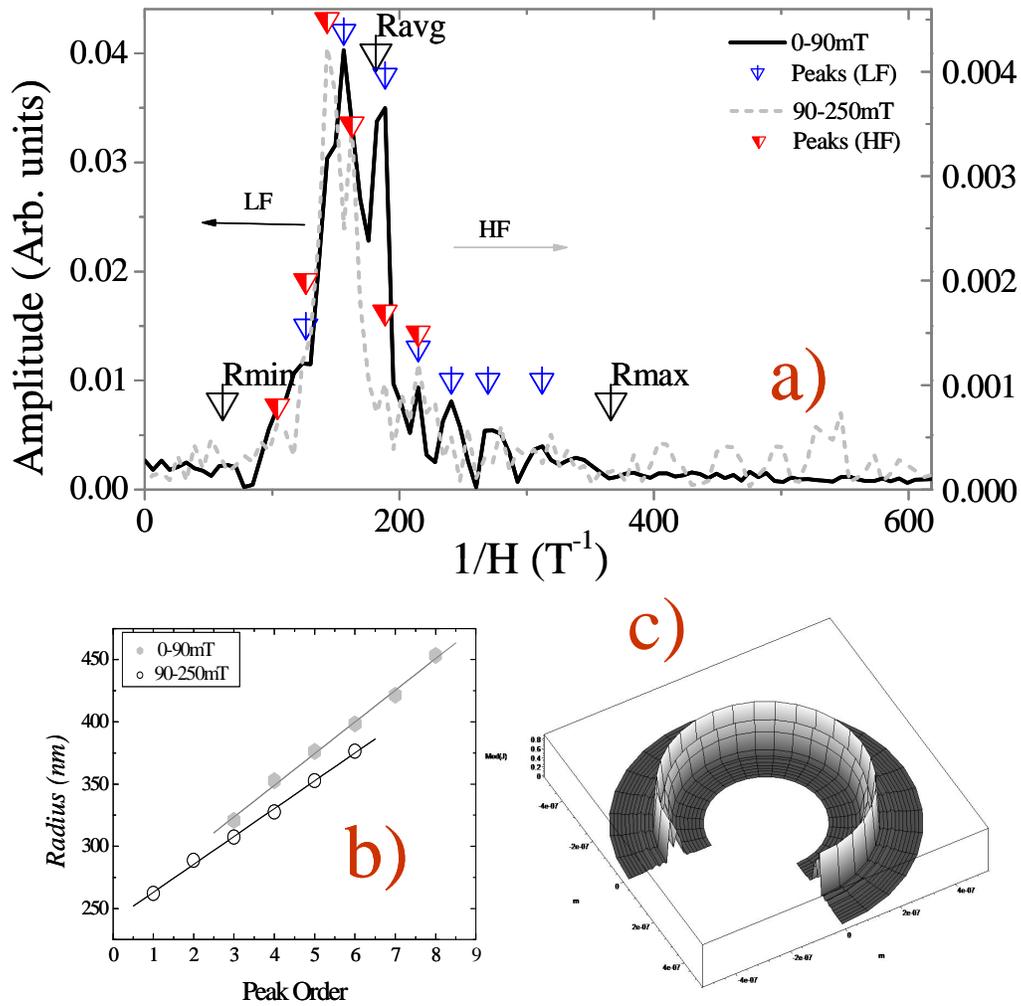

Figure 3 (Color online) a) FFT of data in Figure 2a in the range 0 to 90mT (solid line) and 90 to 250mT (dotted line). From the peaks marked in a) we calculate an effective radius (see text) and plot it as function of peak order (b). In c) we schematize the concentric vortex structure by plotting, in polar coordinate $\rho = (\Phi_0 H/\pi)^{0.5}$ and $\theta$, the amplitude of the FFT in a).